# The effects of marine protected areas over time and species' dispersal potential: A quantitative conservation conflict attempt

**Aristides Moustakas[1,*]**

1. School of Biological and Chemical Sciences

Queen Mary, University of London

Mile End Road, London E1 4NS, UK

\* Corresponding author

Aris Moustakas

Email: arismoustakas@gmail.com

**Abstract**
Protected areas are an important conservation measure. However, there are controversial findings regarding whether closed areas are beneficial for species and habitat conservation as well as for harvesting. Species dispersal is acknowledged as a key factor for the design and impacts of protected areas. A series of agent based models using random diffusion to model fish dispersal were run before and after habitat protection. All results were normalised without the protected habitat in each scenario to detect the relative difference after protecting an area, all else being equal. Model outputs were compared with published data regarding the impacts over time of MPAs on fish biomass. In addition data on species' dispersal potential in terms of km per year are compared with model outputs. Results show that fish landings of species with short dispersal rates will take longer to reach the levels before the Marine Protected Areas (MPAs) were established than landings of species with long dispersal rates. Further the establishment of an MPA generates a higher relative population source within the MPA for species with low dispersal abilities than for species with high dispersal abilities. Results derived here show that there exists a win-win feasible scenario that maximises both fish biomass as well as fish catches.

# 1. Introduction

Habitat protection is a complex issue which has only recently achieved high public visibility [1]. In marine environments it covers many aspects, such as conservation of juvenile fish habitats, protection of corals, and development of marine recreational parks or dive sites. Fishing is often seen as a destructive force, and habitat destruction by fishing practices has to be considered in any comprehensive management plan [2]. Habitat protection can be total or partial. Total closures are often associated with Marine Protected Areas (MPAs) and the designation of certain areas for alternate uses such as recreation.

Closing an area affects several stakeholders: Closed areas are of interest to biologists, conservation scientists, land use planners, but also fishermen and the fishing industry in general as well as the tourism industry [3,4]. While there are cases where closed areas are beneficial for species and habitat conservation [2,5] there are also studies that question the benefits of closures from the economic perspective regarding fish landings [2,6]. This in turn has implications for both food security [7] as well as economic impacts on fisheries [8-10]. Thus a win-win scenario in terms of both increased fish biomass as well as increased fish landings after establishing an MPA is ideal [11] but questionable.

The design of MPAs involves specifying the total surface area to be protected, the distribution in space of that area, and its connectivity [12]. That leaves a fairly wide range of choices: there is controversy about whether single large reserves are more effective than several smaller ones of the same total area (SLOSS), whether edge effects diminish their efficacy, and whether closely spaced reserves are more effective than distantly spaced ones [12]. It is acknowledged that dispersal is a key factor in designing MPAs [13,14]. The reasons behind dispersal been a key factor [15] are that: (i) MPAs should be large enough so that adults can stay long enough inside them, but how large is large enough is clearly related with dispersal potential, (ii) MPAs should be close enough so that larvae can move between them, but how close is close enough is also related to dispersal potential.

Assuming dispersal to be an important factor in determining the ability of species to reach the protected areas then the impacts on species with different dispersal abilities may vary in time since the establishment of an MPA [16,17] for various reasons related with species growth rates or the ability of species to reach or remain within the MPA.. Here, assuming all other factors that influence the efficacy of MPAs to remain equal, it is investigated what are the impacts of MPA(s) on biomass inside and outside the reserves as well as on landings over time to species with different dispersal abilities. In an effort to provide the relative differences in fish biomass and fish landings with and without MPAs, agent based simulation modelling is used [12,18] to model migration [19] via diffusion [20]. Model outputs of each simulation scenario after the establishment of an MPA were normalised by model outputs of the same scenario prior to the establishment of an MPA in order to detect relative changes before and after closing an area.

# 2. Methods
*Model overview and rationale*

A simulation model is used to predict the efficacy of MPAs as a function of species' dispersal potential and different catch rates across two different MPA spatial design scenarios. All results presented here (regarding fish biomass and annual catch) were normalised to 100% in the steady state situation without the MPA in each scenario. Thus, results presented here are presented as dimensionless numbers. Clearly, results from field studies are expected to differ in their values but in comparison with field data the shape of the curves should be at least similar. The model assumes that fish move around at random [21]. Such a modelling attempt can serve as a null model [16] and potentially as a minimal model for pattern formation [22]. This is a conservative (and often an unrealistic) approach as many species exhibit directed dispersal by seasonal migration between feeding and spawning areas. However there are also species that exhibit such dispersal behaviour such as littoral fish species (estuarine fish, intertidal fish, coral reef fish) and the fishery that is mainly involved with this type of fishing is trawl and recreation fishing [23]. In addition, habitat specialist like coral reef species such as clownfish, anemonefish and damselfish are also characterized by this type of movement. The active fishery that is predominantly linked to this type of fish is artisan fishing [24].

*Model description*

The model follows previous modelling attempts where a full description is provided [12,25] modified accordingly here so as dispersal to be random. The model is run on a square grid with 100 x 100 cells and each cell contains a fish biomass value $V(i, j)$. The initial fish biomass concentration was set to $V(i, j) = 100$ for all cells. Time step interval $t$ was set to one day and the total length of the simulation period $T$ was set to 10 years.

Population growth occurs at each time step with a constant (time and space-independent) growth rate $G$. Fish landings (i.e. fish harvesting, thereby Landings, $L$) occur at each time step with a rate of $L$. Landings are distributed over space (cells, $i, j$) at each time step $t$ such that higher fishing mortalities occur at cells with higher fish biomass concentration [26]. Doing so, fishing effort (landings) are proportional to the concentration of fish [27]. Thus, for a given annual mortality rate $M$ fishing mortalities are the same in scenarios with and without MPA(s) but in scenarios that include MPA(s) fish harvesting (in the model landings) is spatially more intensive as the same effort is distributed among fewer cells.

Fish movement is random with an equal probability of diffusing to the eight adjacent neighbouring cells. The probability of migrating to one of the eight neighbouring cells is multiplied by $D$ (dispersal) with values of $D$ close to 0 indicating small dispersal probability and thus a species with short dispersal potential while $D$ values close to 1 indicate long dispersal.

For each time step $t$ and for each cell $i, j$ new biomass $V_{(i, j, t+1)}$ values are updated in all cells prior to the establishment of MPA(s) or in all non-protected cells after MPA(s) are introduced by updating biomass at current cell by adding growth, subtracting natural mortalities and fishing mortalities (landings), adding biomass that potentially diffused from any of the eight neighbouring cells and subtracting biomass from current cell that diffused to only one of the eight neighbouring cells:

$V_{(i, j, t+1)} = V_{(i, j, t)} * (1 + G - (M+L) * V_{(i, j, t)}) + D * ((V_{(i-1, j, t)}$ OR $V_{(i+1, j, t)}$ OR $V_{(i-1, j-1, t)}$ OR $V_{(i+1, j+1, t)}$ OR $V_{(i, j+1, t)}$ OR $V_{(i, j-1, t)}$ OR $V_{(i-1, j+1, t)}$ OR $V_{(i+1, j-1, t)}) - V_{(i, j, t)})$

and *Landings* $= L / [\sum_{i}^{100} V(i,j,t), \text{ for } i,j \in \text{all cells}) - (\sum_{i}^{j} V(i,j,t), \text{ for }, i, j, \in \text{all protected cells})]$

while new biomass $V_{(i, j, t+1)}$ values are updated in all protected cells by:

$V_{(i, j, t+1)} = V_{(i, j, t)} * (1 + G + D * ((V_{(i-1, j, t)} \text{ OR } V_{(i+1, j, t)} \text{ OR } V_{(i-1, j-1, t)} \text{ OR } V_{(i+1, j+1, t)} \text{ OR } V_{(i, j+1, t)} \text{ OR } V_{(i, j-1, t)} \text{ OR } V_{(i-1, j+1, t)} \text{ OR } V_{(i+1, j-1, t)}) - V_{(i, j, t)})$

There are no periodic boundary conditions meaning that fish located in the 4 corner cells of the simulation grid may move only to their three neighbouring cells.

The spatial design of MPA(s) included two different scenarios: a single large and two small MPAs totalling the surface of the single large and in each case the same total surface area was protected. The total surface area protected spanned from 1% up to 20% of the simulation surface area. In all cases mortalities *M* remain constant as prior to the establishment of MPA(s).

In order to examine relative differences with and without MPAs each simulation scenario is replicated with a common parameter space *T, G, M, L, D*: in the first case without an MPA and in the second case with MPA(s). The model assumes that before the imposition of any MPA the fishery dispersing with a dispersal coefficient *D* had reached a steady state with the stock (only one stock is considered) growing at a rate of *G* day$^{-1}$ equal to the natural mortality of *M* day$^{-1}$ (*G = M* in the absence of landings *L*). Thus the fish stock explored exhibits mortality rates *M + L > G*, an over-fished population. Fish biomass *V(i, j, t)* and landings *L(i, j, t)* on cell *i, j* time *t*, are recorded for every cell and time step for each identical simulation scenario (same *T, G, M, L, D*) pre and post MPA(s) establishment and sequentially divided as results post MPA(s) establishment / results pre MPA(s) establishment [$V(i, j, t)_{MPA}/ V(i, j, t)_{noMPA}$ and $L(i, j, t)_{MPA}/ L(i, j, t)_{noMPA}$]. Doing so, the relative change before and after the establishment of MPA(s) is examined.

The simulation scenarios examined here (parameter space) include fish dispersal coefficients *D* varying from 0.1 to 0.2 with increments of 0.02, and from 0.1 to 0.5 with increments of 0.05. Landings were simulated for annual landing rates *L = 1.1\*G*, and *L = 1.25\*G* (landings *L* up to 25% larger than the growth rate *G*). The total surface area protected covered up to 20% of the simulation space. The recorded variables included the development of fish catches over time and the spatial distribution (inside and outside the MPAs) of the simulated stock over time. Each parameter space scenario was replicated 10 times to account for stochasticity and results were averaged.

*Model validation - Confronting model outputs with data*

In order to constrain model outputs with data [18], published data regarding fish biomass of fish species pre and post MPA establishment were used from California Channel Islands, USA, including five fish species (see next paragraph for details regarding species); [28] for model validation. The data included species specific biomass data before and after MPAs establishment [28], allowing comparisons of impacts over time, as well as within and outside the protected area after the MPA was established from 2003 – 2008, allowing comparisons inside and outside protected area after MPAs establishment. Further, the dataset provides also statistics on landings on commercial species before and three years after the establishment of MPAs.

The species specific landings post/landings pre MPAs establishment were regressed against their dispersal potential. Dispersal potential of each species were retrieved from published studies as following: *Semicossyphus pulcher* and *Caulolatilus princeps* from [29]; *Atractoscion nobilis* from [30]; *Ophiodon elongatus* from [31]; *Paralicthys californicus* from [32]. Further in order to investigate the ratio of fish biomass inside and outside MPAs after the establishment of MPAs, density (number of fish per 100 m$^2$) of three targeted fish species was retrieved at the same time snapshot inside and outside MPAs and regressed against the species' dispersal potential. The three fish species included *Semicossyphus pulcher*, *Ophiodon elongatus*, and *Sebastes miniatus* and their dispersal potential was retrieved for the first two species as cited above. Dispersal potential of *Sebastes miniatus* was retrieved from [33].

In order to link model predictions with marine species dispersal potential and thus predict the time impacts on landings of different species groups, analysis on (adult) marine taxa dispersal data was conducted. The data derived from a meta-analysis of 1897 publications [34,35]: Within this dataset a search regarding dispersal rate of species was conducted. From the 1897 publications, only the ones that explicitly mentioned dispersal rates per species and length of the study so that dispersal can be normalised as km year$^{-1}$ were used. In total the dispersal rates of N = 553 marine taxa were available in the dataset.

An Empirical Cumulative Density Function (ECDF) was used to evaluate the dispersal range of each species (in km) against the percentage of species in the dataset that have a dispersal potential less than or equal to that value. The ECDF $F_n(x)$ is defined as:

$$F_n(x) = \frac{number\ of\ elements\ in\ the\ sample\ \leq x}{n} = \frac{1}{n}\sum_{i=1}^{n} 1\{t_i \leq x\}$$

In the case examined here the values of $F_n(x)$ on the vertical axis define the percentage of all species ($t_i$) with a dispersal range less than or equal to the corresponding value on the horizontal axis, $x$ in km yr$^{-1}$. For example the value on the vertical axis of 10 corresponds to the value x on the horizontal axis of the dispersal range in km yr$^{-1}$ of 10% of all species. The ECDF resembles a cumulative histogram without bars and it is based on parameters estimated from the original data [36]. In this respect, an ECDF is similar to a probability plot except both axes are linear and non-transformed [36]. Further, 95% confidence intervals of the mean and median values of species' dispersal rates were calculated.

## 3. Results

Model outputs showed that recover of landings (in comparison to the levels of pre-MPA establishment) was faster for species with high dispersal rates than for low dispersal rates and this applies to both single large and several small MPAs spatial design for mortalities ($M+ L$) up to 25 % larger than growth rates $G$ and for 20% of the total surface areas been protected (Fig. 1a; 2a). This implies that landings of species with low dispersal rates or short home rates will take longer to recover. Spatial distribution of species biomass within the MPA(s) increases with decreasing dispersal potential and this applies to both single large and several small MPAs for mortalities up to 25 % larger than growth rates and for 20% of protected surface area (Fig. 1b; 2b). Results for mortalities *M < 1.25\*G* produced higher recovery of landings and biomass (results not shown here). However, results for total protected surface area < 20% resulted in the recoveries of species with high dispersal rates only (results not shown here).

Statistical analyses of fish density data post and pre MPA establishment showed that landings of commercial fish species in post MPAs establishment divided by the landings of the same species pre MPAs establishment regressed against the dispersal potential of each species showed that in the case of 5 commercial fish species examined the relative change in landings post normalized by pre MPAs establishment was more pronounced in species with longer dispersal rates (Fig. 3a; $R^2$ = 70.8%, df = 1, p = 0.022 formula: $\log_{10}$(after/before) = - 0.4725 + 0.1860 $\log_{10}$(Dispersal). Relative fish density inside divided by fish density outside MPAs regressed against the species' dispersal potential showed that species with shorter dispersal rates have relatively shorter density inside than outside MPAs (Fig. 3b; $R^2$ = 100%, df = 1, p=0.008, formula: $\log_{10}$(fish density in/out from the MPA) = 0.2785 - 0.05567 $\log_{10}$(Dispersal).

Highest dispersal potential is exhibited amongst the phyla of Gadirormes, Crustaceans, Perciformes, Echinodermata, Mollusca, and Pleuronectiformes (Fig. 4a). With the exception of Gadirormes and Pleuronectiformes, phyla with high dispersal rates, have high variation of dispersal rates between individual species within the phylum (Fig. 4a). The majority of phyla examined have dispersal rates of less than 1 km (Fig. 4a). From the species considered here, 48% of all species have dispersal rates of < 1 km, while 90 % of all species have dispersal rates of < 200 km (Fig. 4b). Overall, dispersal rates between species was very high as indicated by differences between 95% confidence intervals of the mean = 54 km, [41, 68] and the median = 7 km, [4, 31].

## 4. Discussion

Model outputs derived here depict the relative time needed for fish landings to reach levels before the establishment of an MPA. The method - normalizing outputs after a change in the system has been introduced by model outputs prior to the change - may serve as a valuable null model tool in ecology and biological sciences in order to investigate the relative effects of a key parameter (here dispersal on the impacts of MPAs on both fish biomass and landings). Models are used when experiments are costly, require significant labour effort, ethics, and effects of spatial or temporal scales associated. Cellular automata and agent based models are useful tools for addressing such issues [12,25,37-41].

*Recovery after the establishment of an MPA as a function of dispersal*

Model outputs derived here showed that fish catches are more likely to recover faster at the original levels pre-MPA(s) establishment and above. Statistical analysis of normalized post/pre MPAs establishment data exhibited a monotonic pattern, faster recovery of landings of long dispersers - data were available for five species and five years after closures. Previous spatially-explicit studies on population recovery after disturbance have indicated that long-dispersers recover more homogeneously than short-dispersers [42,43], and to that end model outputs are in agreement with this. For additional discussion on the interplay between highly mobile fish and the efficacy of MPAs see also [44].

*Source - sink dynamics and biomass inside & outside MPAs*

Source-sink theory has been applied to the spatial design and impacts of MPAs [45,46]. Results derived here exhibited that MPAs are increasingly acting as population sources as species' dispersal range decreases. Species with shorter dispersal rates are likely to be also smaller in size and/or body mass [47,48] and thus they benefit more simply by the fact that in all scenarios MPAs had an equal total size. Clearly, home range areas of short dispersers will be smaller than those of long distance dispersers (the model does not account for individual's body length or mass). However, given that species with short dispersal potential have more restricted distributions [49,50] overall it seems reasonable to expect that protecting the habitats of short dispersers will create larger population buffers within the protected area than when protecting the habitats of long dispersers. Data of movement of lingcod (*Ophiodon elongates*, a species with limited dispersal rates, from the 5 examined species post/pre MPAs landings) in and out of an area closed to fishing showed that individuals left the reserve but were only absent for short time [31]. Model outputs from another study have also reported that modest dispersal rates of fish can reduce abundance within protected areas [51].

According to the results derived here, the abundance of species of phyla with very low dispersal rates such as Porifera, Rhodophyta, Bryozoa, and Anthophyta will be considerably higher within the MPA than outside. The majority of these species are not commercial (and thus would not be targeted by fishers) but a 'blind' fishing method such as trawling would affect them [52,53]. Further, several of the short dispersing species are habitat forming species [54]. It should be noted however, that these conclusions are based upon a fairly large dataset [34,35], but this dataset is not exhaustive.

In general the variables used in this work have no units, as they are normalised. However when comparisons with real fish species is made, since real *D* values are used, it would be interesting to know them, and gain an insight of the real-life size of the grid, and subsequent grid-cell size, and MPA size used. The model is run on a simulation space of 100 x 100 = 10,000 cells. Assuming a perfectly directed dispersal (the opposite of random diffusion) from the upper left to the lower right corner of the simulation grid, which is the maximum straight line distance that can be made, fish can disperse 141.42 cells which is the diagonal. The minimum value of dispersal recorded in the dataset was 0.0005 km yr$^{-1}$, while the maximum was 527 km yr$^{-1}$. Defining the diagonal distance by the largest dispersal value then 142 cells correspond to 527 km and thus the cell diagonal is ~3.7 km, the cell side ~2.6 km, the cell surface area ~6.9 km$^2$ and the simulated area ~6.8 x 10$^4$ km$^2$. Note that these values are only listed as a gross rule of thumb as (a) species disperse randomly and not directed, and (b) the ABM model is not scale-specific calibrated [18,55].

<u>Limitations and simplifications of the method</u>

This study shows that in the parameter space explored a win-win scenario in terms of fish biomass and landings increase after some years of closing an area is feasible; but it does not show what the actual parameter space leading to this result is, it only shows that mathematically this is possible. Despite the fact that the results presented here are unit-less (ratio) the sensitivity to the scale of analysis has not been accounted for [56] in terms of multi-scale modelling [57]: A ratio is scale-free, but the actual processes as they are defined here are not. There are several scales involved: D in the context of a diffusion process regards dispersal distance squared divided by time and thus, both space and time scale is involved [58]. Due to the implicit scale of the grid cells (unit size) relative to unit time increment D is a dimensionless number in the present model. If these cells had been defined smaller, D would have to be increased to maintain the present results for the given parameters and within the given MPA(s) and total area. Thus, the crucial aspect for the present results is the dispersal rate relative to refuge size [59]. In addition the model population has no age structure and there is no density dependent regulation [60], though a study over 14 y showed that density dependence was still not halting development of the population within the MPA [61]. With respect to the latter, growth rate is set constant both in absence and presence of fishing and mortality from fishing is also set proportional with fish biomass in unprotected areas. Consequently, density is in the MPAs assumed to be below carrying capacity [62,63]. The lack of age-structured population dynamics may be defended for species with a natal dispersal rate that is smaller than adult dispersal rate [64]. Otherwise, population renewal is not sufficiently concentrated inside the MPAs to achieve the observed source/sink results as it happens in reality. This is also evident in simulation outputs inside MPAs, since densities under low D may reach more than tenfold increase relative to pre-MPA levels.

The time interval of simulation (10 years) may seem short, because the effect of MPAs is usually visible after long time intervals [17] and the lifespan of some species may exceed this time. Moreover, Fig.1a and Fig. 2a suggest that with a longer time interval more curves could reach the 100% target. However, in general there are several behavioural changes in fishers after establishing an MPA [65]. While it would be interesting to know whether landings attain the levels observed before the implementation of MPAs (convex curves for high dispersal distance) and how long this will take, other acting processes such as increased fishing pressure [66] or phenotypic evolution [67,68] are also acting and thus long-term outputs are unlikely to be realistic. Thus, the model was only run long enough to discern some variability between species' dispersal abilities.

In addition in the simulation grid, the corner cells get inputs only from their three neighbouring cells, giving a lower growth at the edge of the area as no periodic boundary conditions were used For a view on scaling issues in gridded models and model structure with scenario boundary conditions see also discussion in [18,69]

*Conclusions*

There are very large differences in dispersal potential of species as indicated also by differences between mean (~ 50 km yr$^{-1}$) and median (~ 7 km yr$^{-1}$); [70]. The mean dispersal value is derived mainly by relatively few species with long dispersal potential. The median dispersal value is rather reflecting the dispersal potential of the majority of species. In addition the ECDF distribution values indicate that 50% of all species disperse no more than 1 km per year and 70% of all species no more than 50 km yr$^{-1}$. Distances between MPAs often are not comparable to these values [45]. This indicates that there is no single optimal conservation strategy if dispersal is a critical factor for the efficacy of MPAs: Large-bodied marine species are under greater threat of global extinction [71]. Large-bodied species have longer dispersal rates [49]. It is therefore difficult to design an MPA that will account for long dispersers and thus large-bodied threatened species, and simultaneously account for maximizing biodiversity within the MPA (based on dispersal as a biodiversity proxy) or maximize slow-dispersing habitat building species.

Introducing MPAs may lead to a temporary decline of landings, owing to stronger fishing effort outside the protected areas to compensate for lack of fishing inside MPAs. However, over time the source/sink effect – due to a gradual many-fold increase in fish abundance inside the MPAs – may not only gradually make landings from the unprotected fishing areas rising again but even overshoot the pre-MPA level. This result was achieved under over-fishing, a 25% of total mortalities (natural and fishing mortalities) higher than the growth rate as it often happens in reality [72,73]. Thus, a win-win result is achieved [11]: fish and local ecosystem is protected and thriving inside protected areas, and fishery will benefit from a net gain after a temporary decline while waiting for the MPA population(s) to increase sufficiently to become a strong provider of dispersing individuals [11]. This win-win scenario needs time [11,74] and in general an integration of science and stakeholder based methods may facilitate such scenarios [75,76].

Fast recovery or even overshoot of landings relative to pre-MPA level basically depends – under the given model design – on two main aspects: Dispersal rate D and number of MPAs (actually, the size of MPAs relative to D; see below). Larger D and/or splitting of MPA into a set of smaller areas with the same total area both contribute positively to reducing time to regaining pre-MPA landing quantity. Thus, in the context of SLOSS, from the present results many small refuges seem to benefit both fish populations and exploitation. D is species- and habitat dependent (and also varying with age class, which is not accounted for here). However, number of MPAs, their locations and sizes are manageable. This theme has been subject to much research, both empirically and by simulations, and results from meta-analyses have generally been non-conclusive due to the many-faceted system dynamics [77]. Theoretical results have generally supported the a priori intuitive hypothesis that strong dispersers are less protected by MPAs than more sedentary species [12,25,78]. However the present results support the opposite: strong D leads to relatively fast recovery of landings after implementation of the refuge, while still maintaining larger fish density inside the refuges relative to pre-MPA level. Splitting the refuge into smaller entities improves recovery even better, and may lead to even better fishing yield in the long run.

At present MPAs are generally covering much less than 20% of the fishing scale extent, and consequently this policy need revision in order to achieve the net fishing gain over time. Other studies suggested that that the yield from the harvest effort is strongly affected by the fraction of area protected from harvesting and that maximum yield is independent of the size of the protected area unless the fraction is > 0.56 [59] The dependence on D is a key parameter here, and should be considered relative to (dispersal distance squared)/(time unit) and MPA size, and an estimate of fishing range in the actual area.

**Acknowledgments**
Comments of two anonymous reviewers and the handling editor Ronald Brandl considerably improved an earlier manuscript draft. This paper is dedicated to William (Bill) Silvert with whom I very much would have liked to write the paper together.

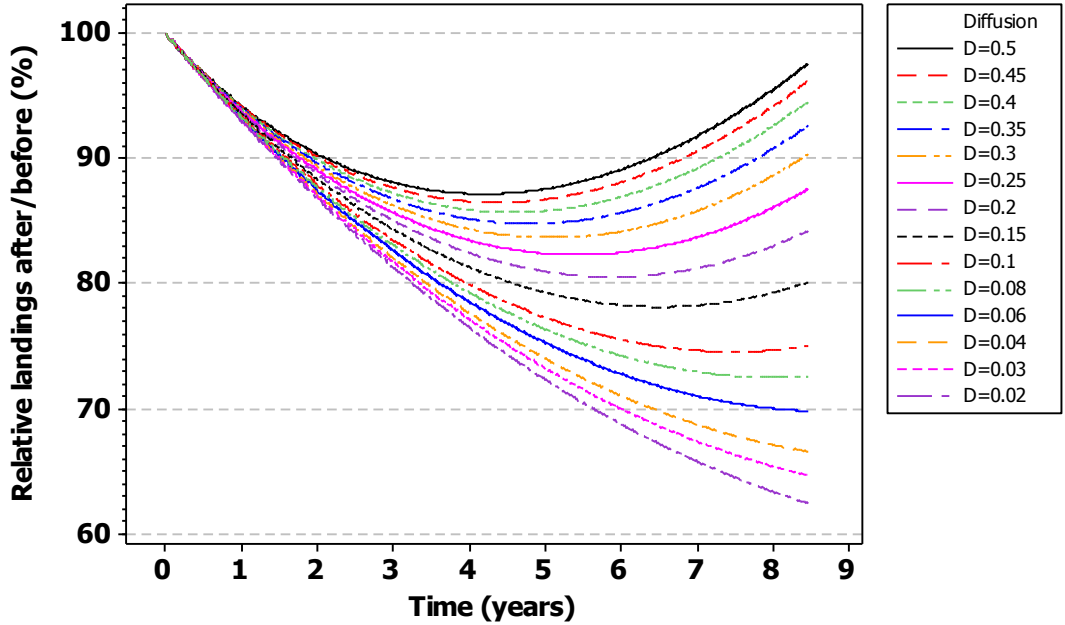

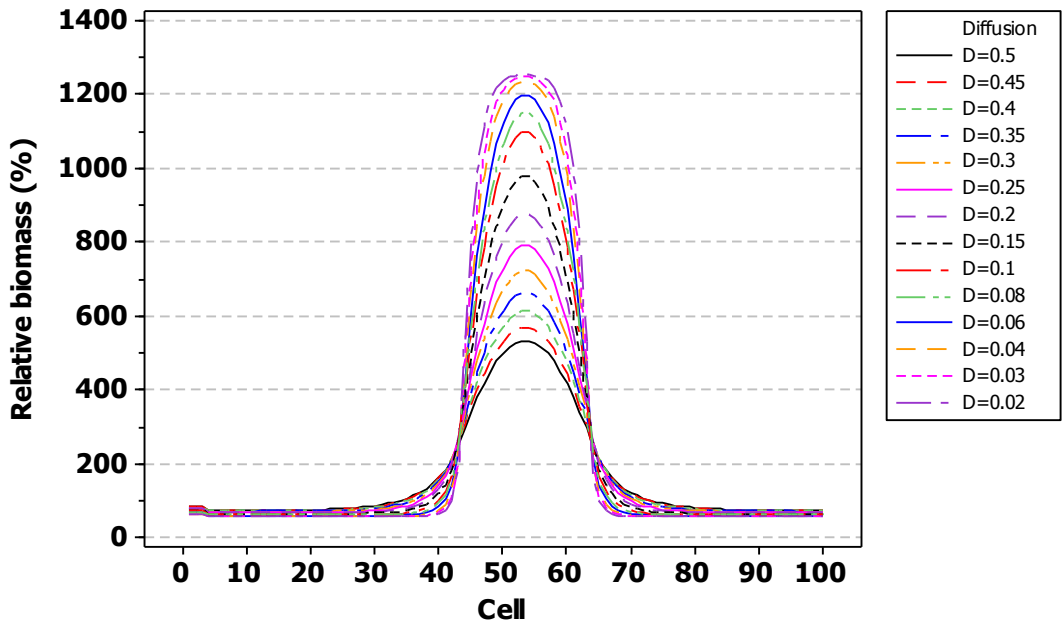

**Figure 1. a.** Post-MPA establishment landings over time (x-axis) normalised by the landings before the establishment of an MPA, replicated for a number of species' dispersal potential (diffusion with values *D*). **b.** Post-MPA establishment spatial distribution of biomass normalised by the one prior to the MPA establishment. All results are referring at outputs at the end of simulation period (10th year). All simulation scenarios assume that 20% of the cells are protected with a single MPA. All results presented here (regarding fish biomass and annual catch) were normalised to 100 in the steady state situation without the MPA in each scenario and thus, are unit-less numbers.

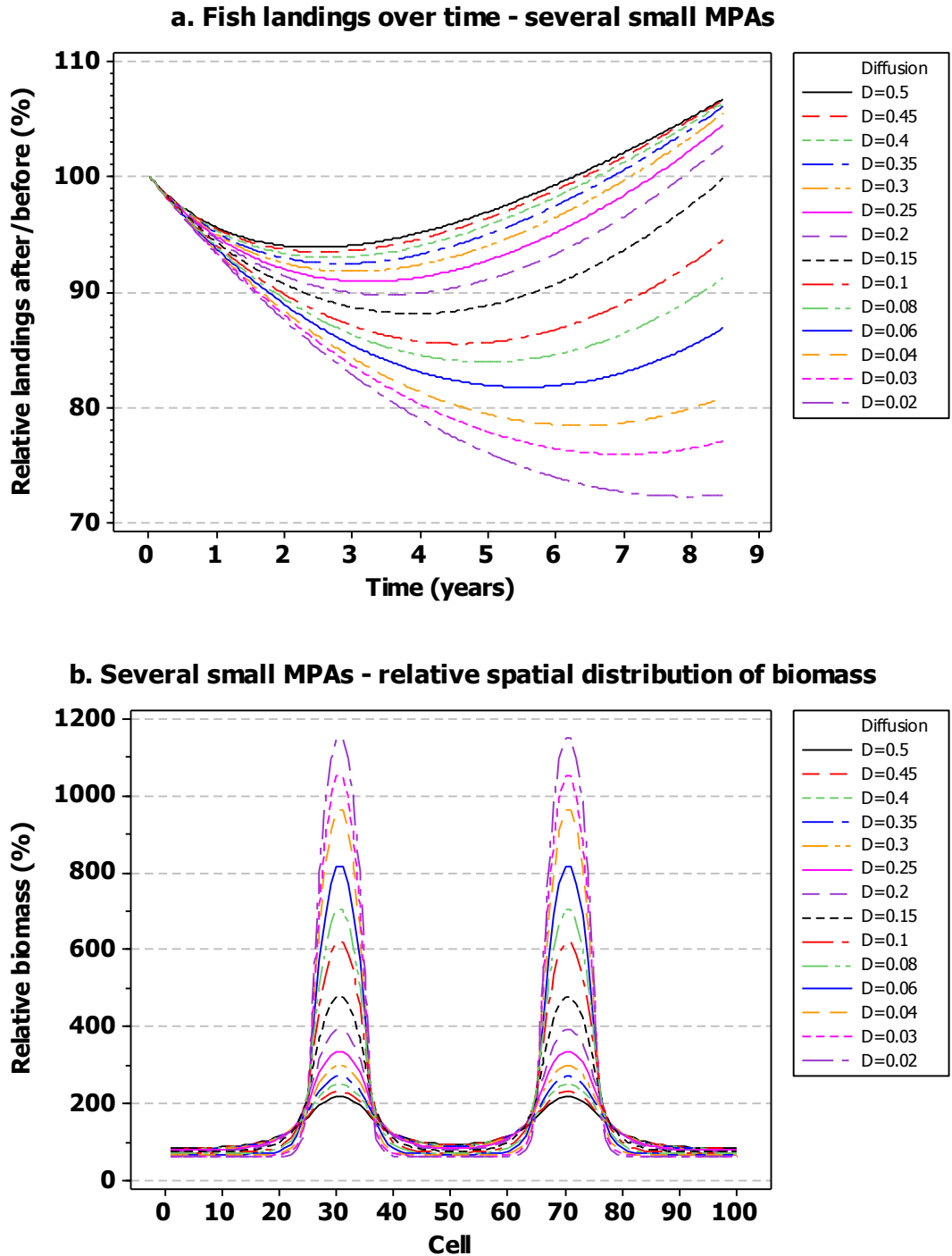

**Figure 2. a.** Post-MPAs establishment landings over time (x-axis) normalised by the landings before the establishment of two MPAs, replicated for a number of species' dispersal potential (diffusion with values *D*). **b.** Post-MPAs establishment spatial distribution of biomass normalised by the one prior to the MPAs establishment. All results are referring at outputs at the end of simulation period (10th year). All simulation scenarios assume that 20% of cells are protected with two equal-sized MPAs. All results presented here (regarding fish biomass and annual catch) were normalised to 100 in the steady state situation without the MPAs in each scenario and thus, are unit-less numbers.

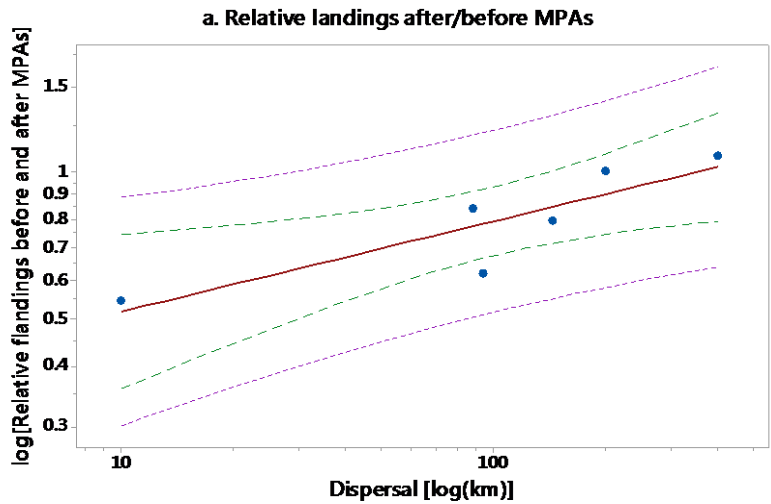

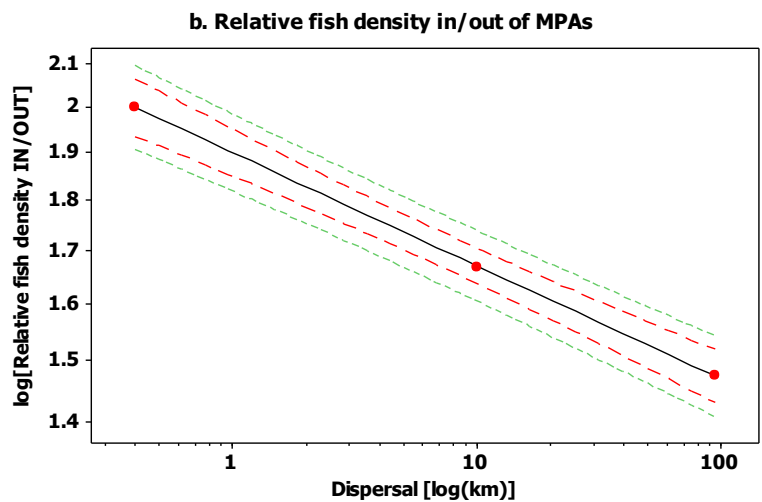

**Figure 3a.** Data of landings of 5 commercial fish species post MPAs establishment [28] normalised by the landings of the same species pre MPAs establishment . The ratio of post/pre MPAs establishment landings was regressed against the dispersal range of each species. **3b.** Data of relative fish density inside / outside the MPAs. The ratio of in/out MPAs relative fish density was regressed against the dispersal range of each species (see section 'Confronting model outputs with data' for more details). Solid lines are the best fit regression, dashed lines the 95% confidence interval, and the dotted lines the 95% predicted interval.

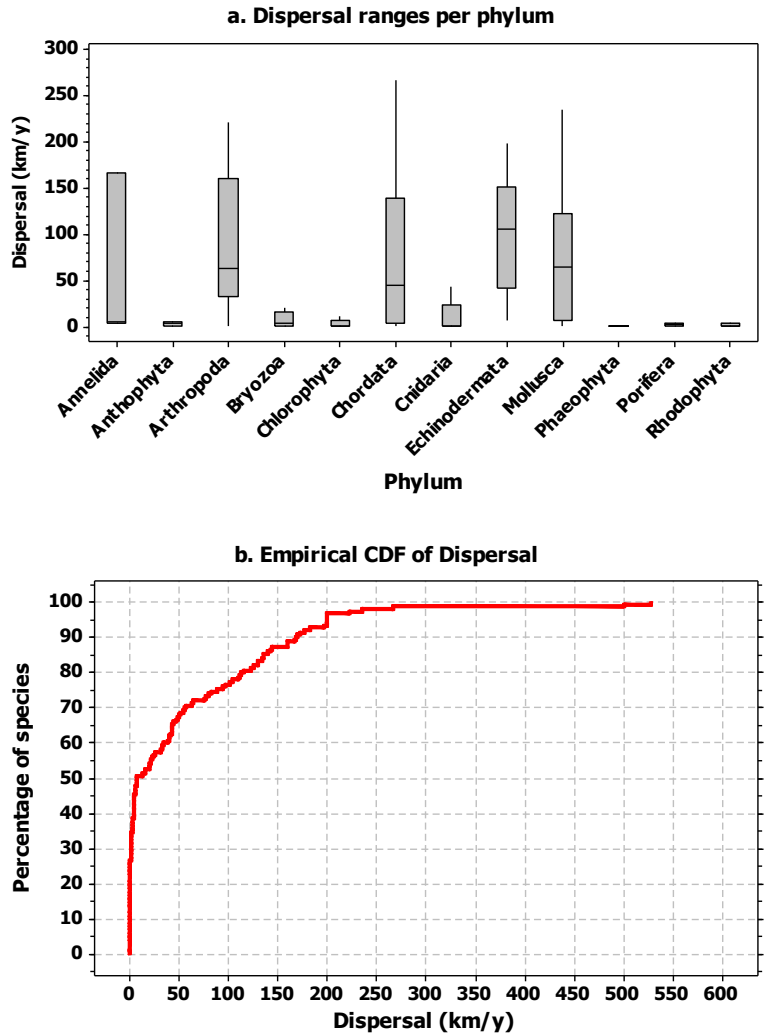

**Figure 4. a.** Dispersal rates per different orders of species (in km year$^{-1}$), data from a search in the dataset described in [34,35]. The solid line is the median, and the boxes are defined by the upper and lower quartile (25th and 75th percentiles). The whiskers extend up to 1.5 times the inter-quartile range of the data. **b.** Empirical Cumulative Density Function (ECDF) of the dispersal range of each species (in km) against the percentage of species in the dataset that have a dispersal potential less than or equal to that value. ECDF shows the percentage of species that exhibit a dispersal range (km yr$^{-1}$) less than or equal to the value on the horizontal axis.